\documentclass{emulateapj}

\newcommand{\msun}{M$_{\sun}$}

\newcommand{\ldl}{$\lambda/{\Delta}{\lambda}$}

\newcommand{\teff}{T$_{\rm eff}$}

\newcommand{\ammon}{NH$_3$}
\newcommand{\meth}{CH$_4$}
\newcommand{\wat}{H$_2$O}

\newcommand{\kms}{km~s$^{-1}$}

\newcommand{\name}{WISEP~J045853.90+643452.6}
\newcommand{\namesh}{WISE~J0458+6434}

\slugcomment{Submitted to ApJ 27 August 2011, Accepted 18 October 2011}

\shorttitle{Resolved Spectroscopy of WISE~J0458+6434AB}
\shortauthors{Burgasser et al.}

\begin{document}

\title{Resolved Spectroscopy of a Brown Dwarf Binary at the T Dwarf/Y Dwarf Transition\footnote{Data presented herein were obtained at the W.\ M.\ Keck Observatory, which is operated as a scientific partnership among the California Institute of Technology, the University of California and the National Aeronautics and Space Administration, and made possible by the generous financial support of the W.\ M.\ Keck Foundation.}}

\author{
Adam J.\ Burgasser\altaffilmark{a},
Christopher R.\ Gelino\altaffilmark{b},
Michael. C.\ Cushing\altaffilmark{c},
and
J.\ Davy Kirkpatrick\altaffilmark{b}
}

\altaffiltext{a}{Center for Astrophysics and Space Science, University of California San Diego, La Jolla, CA 92093, USA; aburgasser@ucsd.edu}
\altaffiltext{b}{Infrared Processing and Analysis Center, MC 100-22, California Institute of Technology, Pasadena, CA 91125, USA}
\altaffiltext{c}{Ritter Astrophysical Research Center, University of Toledo, MS 113, Toledo, OH 43606, USA}

\begin{abstract}
We report resolved near-infrared imaging and spectroscopic observations of the T8.5 binary WISEP~J045853.90+643452.6AB
obtained with Keck/NIRC2, Keck/OSIRIS and the Keck Laser Guide Star Adaptive Optics system.  These data confirm common proper and radial motion for the two components, and we see the first indications of orbital motion (mostly radial) for this system.
$H$-band spectroscopy identifies both components as very late-type brown dwarfs with strong H$_2$O and CH$_4$ absorption. The spectrum of WISE~J0458+6434B also exhibits a compelling signature of NH$_3$ absorption over 1.52--1.54~$\micron$ when compared to the T9 dwarf UGPS~J072227.51$-$054031.2.  Comparison to T8-Y0 spectral standards and $H$-band spectral indices indicate classifications of T8.5 and T9.5 for these two components, approaching the boundary between the T dwarf and Y dwarf spectral classes.  
\end{abstract}


\section{Introduction}

The lowest-temperature brown dwarfs--the L, T and Y dwarfs--span an astrophysically-rich parameter space of mass, temperature and atmospheric chemistry that bridges the lowest-mass stars and the warmest exoplanets (\citealt{2005ARA&A..43..195K} and references therein).  
These sources also probe the efficiency of low-mass star formation and the chemical enrichment history of the Galaxy \citep{2004ApJS..155..191B, 2008ApJ...676.1281M, 2009MNRAS.392..590B}.  Very low-temperature brown dwarfs down to effective temperatures ({\teff}s) of $\sim$300--500~K  are now being identified (e.g., \citealt{2010MNRAS.408L..56L,2011ApJ...730L...9L,2011arXiv1103.0014L,2011arXiv1108.4678C,2011arXiv1108.4677K}), primarily in deep, wide-field infrared surveys such as 
the United Kingdom Infrared Telescope Deep Sky Survey (UKIDSS; \citealt{2007MNRAS.379.1599L}); 
the Canada-France Brown Dwarf Survey (CFBDS; \citealt{2008A&A...484..469D})
and, most recently, the Wide-field Infrared Survey Explorer (WISE; \citealt{2010AJ....140.1868W}).
These discoveries probe previously unexplored realms of  gas and condensate
atmospheric chemistry \citep{2003ApJ...596..587B,2006asup.book....1L}, and encompass the newly identified Y dwarf spectral class \citep{2011arXiv1108.4678C}.

While changes in spectral morphology are largely tied to changes in temperature, the transitions between late-type spectral classes also trace significant changes in atmospheric chemistry and dynamics.  The M dwarf to L dwarf transition is accompanied by the formation of mineral condensates  \citep{1996A&A...308L..29T,2003ApJ...586.1320C,2008ApJ...675L.105H}, which deplete the photosphere of metal oxide gases (TiO, VO) and lead to more reddened optical and near-infrared spectral energy distributions.  In addition, magnetic coronal and chromospheric emission drops precipitously across this M/L transition \citep{2000AJ....120.1085G,2004AJ....128..426W}, the result of both increasingly neutral atmospheres \citep{2002ApJ...577..433G,2002ApJ...571..469M} and reduced magnetic field energy with reduced mass \citep{2009Natur.457..167C}.
The L dwarf to T dwarf transition, signaled by the emergence of {\meth} absorption at near-infrared wavelengths, is coincident with the removal of mineral condensate clouds from the photosphere \citep{2001ApJ...556..872A,2002ApJ...571L.151B,2004AJ....127.3553K,2005ApJ...621.1033T} and the increased influence of non-equilibrium chemistry on observed molecular abundances \citep{1997ApJ...489L..87N,1999ApJ...519L..85G,2006ApJ...647..552S,2010ApJ...722..682Y}.   Given the very recent discovery of Y dwarfs, transitional behaviors have yet to be explored at these spectral types, although atmospheric models predict the formation of salt and water condensates that may significantly shift photospheric chemistry and spectral appearance (e.g., \citealt{1999ApJ...519..793L,2003ApJ...596..587B,2007ApJ...667..537L}). 

Ideal probes of such spectral transitions are physical binaries whose components straddle the corresponding classes.  These presumably coeval systems share a common age and bulk chemical composition, and reside at a common distance, making relative comparisons less dependent on these parameters.  For example, evidence for a rapid dissipation of clouds at the L dwarf/T dwarf transition includes the apparent brightening of early T dwarfs at 1~$\micron$ as observed among L/T transition binaries \citep{2006ApJS..166..585B,2006ApJ...647.1393L,2008ApJ...685.1183L}, and the higher apparent multiplicity rate of these systems \citep{2007ApJ...659..655B}.  Physical parameters such as age, mass and radii have also been measured for low-mass binaries; these serve as critical empirical constraints for evolutionary models (e.g., \citealt{2001ApJ...560..390L,2004ApJ...615..958Z,2006Natur.440..311S,2009ApJ...692..729D,2010ApJ...711.1087K}). Finally, companions to low-luminosity sources are frequently low-temperature extrema.
The prototypes for both the L dwarf (GD 165B; \citealt{1988Natur.336..656B}) and T dwarf (Gliese~229B; \citealt{1995Natur.378..463N}) spectral classes were identified as companions to low-luminosity stars, as were two of the coldest brown dwarfs currently known, WD 0806-661B \citep{2011ApJ...730L...9L} and CFBDSIR~J1458+1013B \citep{2011arXiv1103.0014L}.

One of the first very late-type T dwarfs to be uncovered by the WISE survey was {\name} (hereafter {\namesh}; \citealt{2011ApJ...726...30M}), a T8.5 dwarf at an estimated distance of 10~pc from the Sun.  Subsequent adaptive optics (AO) imaging observations by \citet{2011AJ....142...57G} revealed this source to be a binary with a separation of 0$\farcs$5 and a relative brightness of $\Delta{J}$ = 0.98$\pm$0.08.  Given the late spectral type of the composite spectrum and the significant magnitude difference between the components, {\namesh}B was estimated to have a T9 spectral type and {\teff} $\approx$ 500~K.  However,  uncertainties in the distance of this system, the unknown spectral energy distribution of the secondary, and the very recent empirical definition of the Y dwarf spectral class \citep{2011arXiv1108.4678C} means that {\namesh}AB could potentially straddle the T dwarf/Y dwarf transition.

In this article, we present resolved near-infrared imaging and spectroscopy of the {\namesh} system obtained with the 
Keck II NIRC2 camera, OH-Suppressing InfraRed Integral field Spectrograph (OSIRIS; \citealt{2006SPIE.6269E..42L}) and Laser Guide Star Adaptive Optics system (LGSAO; \citealt{2006PASP..118..297W, 2006PASP..118..310V}).
These data confirm common proper motion for the system and the first evidence for orbital motion.  We use the $H$-band OSIRIS spectra to classify the components, and identify tentative evidence for near-infrared {\ammon} absorption in {\namesh}B.
In Section~2 we describe our observations and data reduction methods.
In Section~3 we analyze our astrometric measurements, and provide an initial qualitative characterization of the binary orbit.
In Section~4 we analyze the spectral data, deriving classifications 
using both spectral templates and $H$-band indices, and examining evidence for {\ammon} absorption.
In Section~5 we discuss our results, focusing on the absolute magnitudes of {\namesh}AB in the context of the T dwarf/Y dwarf transition.

\section{Observations}

\subsection{NIRC2 Imaging}

New high-resolution LGSAO imaging observations of WISE J0458+6434 were obtained
with NIRC2 on the 10 m Keck II Telescope on 2011 August 29 (UT). Conditions were
clear with seeing of 0$\farcs$5. Data were obtained with the narrow plate scale
mode of NIRC2 (9.963$\pm$0.011 mas pixel$^{-1}$; \citealt{2006ApJ...649..389P}) at an orientation of 0$\degr$ and airmass $\approx$1.5.  We
observed the target in the $J$-, $H$- and $K_s$-band filters using
a 3-point dither pattern that avoided the noisy, lower left quadrant of
the focal plane array.   Individual exposure times were 120~s in $J$ and $K_s$ and 300~s in $H$, for total integrations  of 360~s, 900~s and 360~s in $J$, $H$ and $K_s$, respectively. 
The sodium LGS provided the wave front reference source for AO correction, while tip-tilt aberrations and quasi-static changes were measured by monitoring the $R$ = 15.6 field star USNO-B1.0 1545-0122611 \citep{2003AJ....125..984M} located $\rho$ = 27$\farcs$7 from {\namesh}. 

\begin{figure*}
\epsscale{0.9}
\centering
\includegraphics[height=0.45\textwidth]{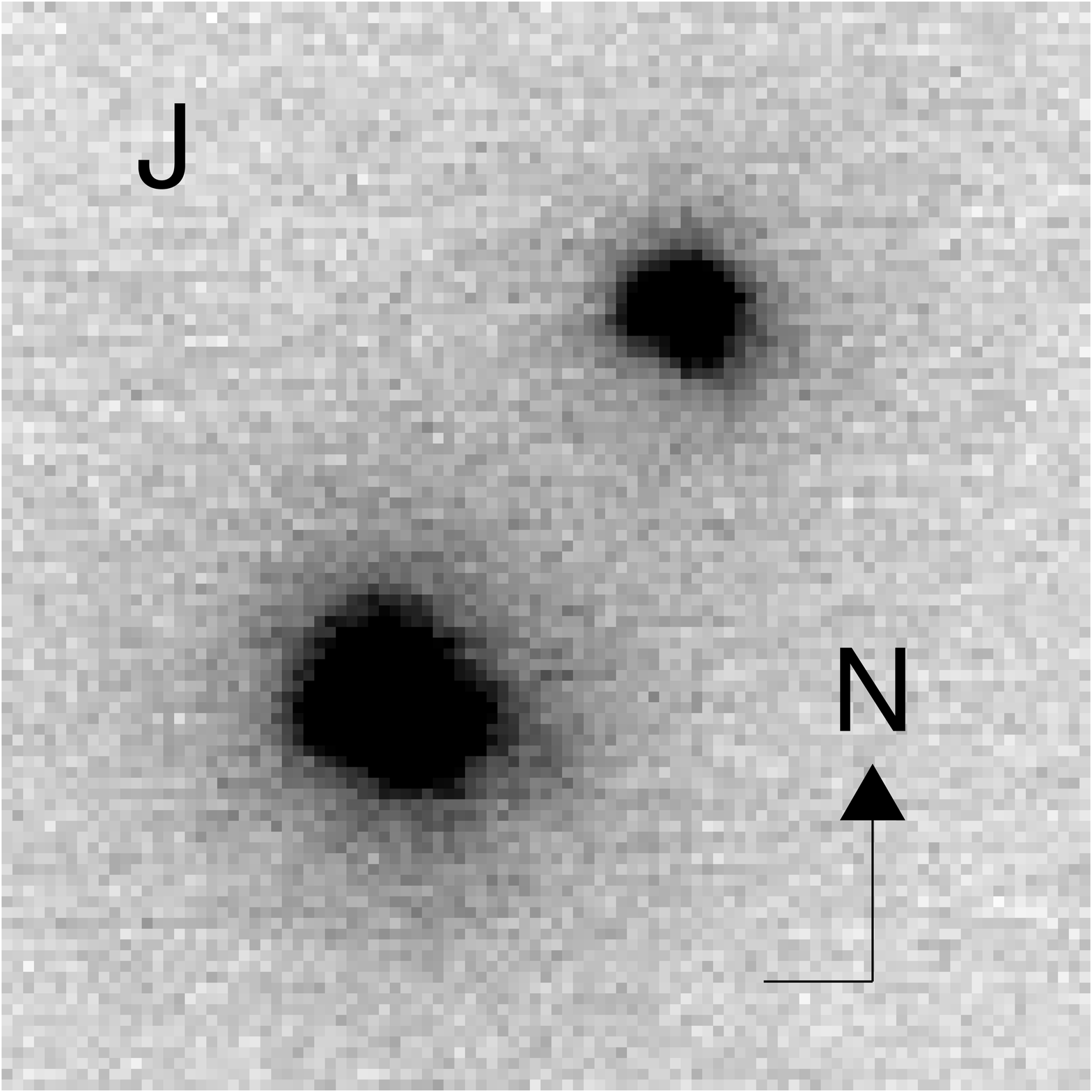}
\includegraphics[height=0.45\textwidth]{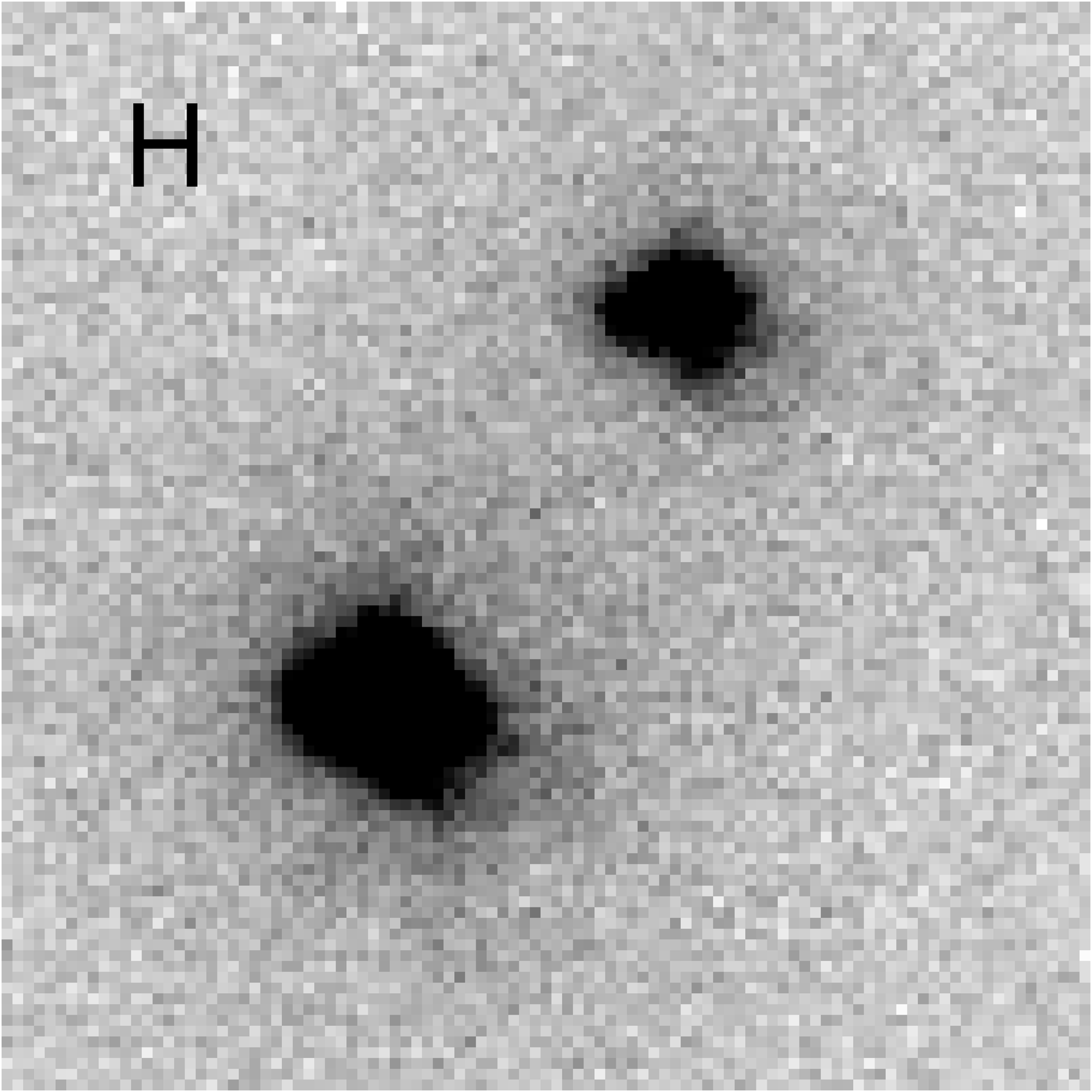}
\includegraphics[height=0.45\textwidth]{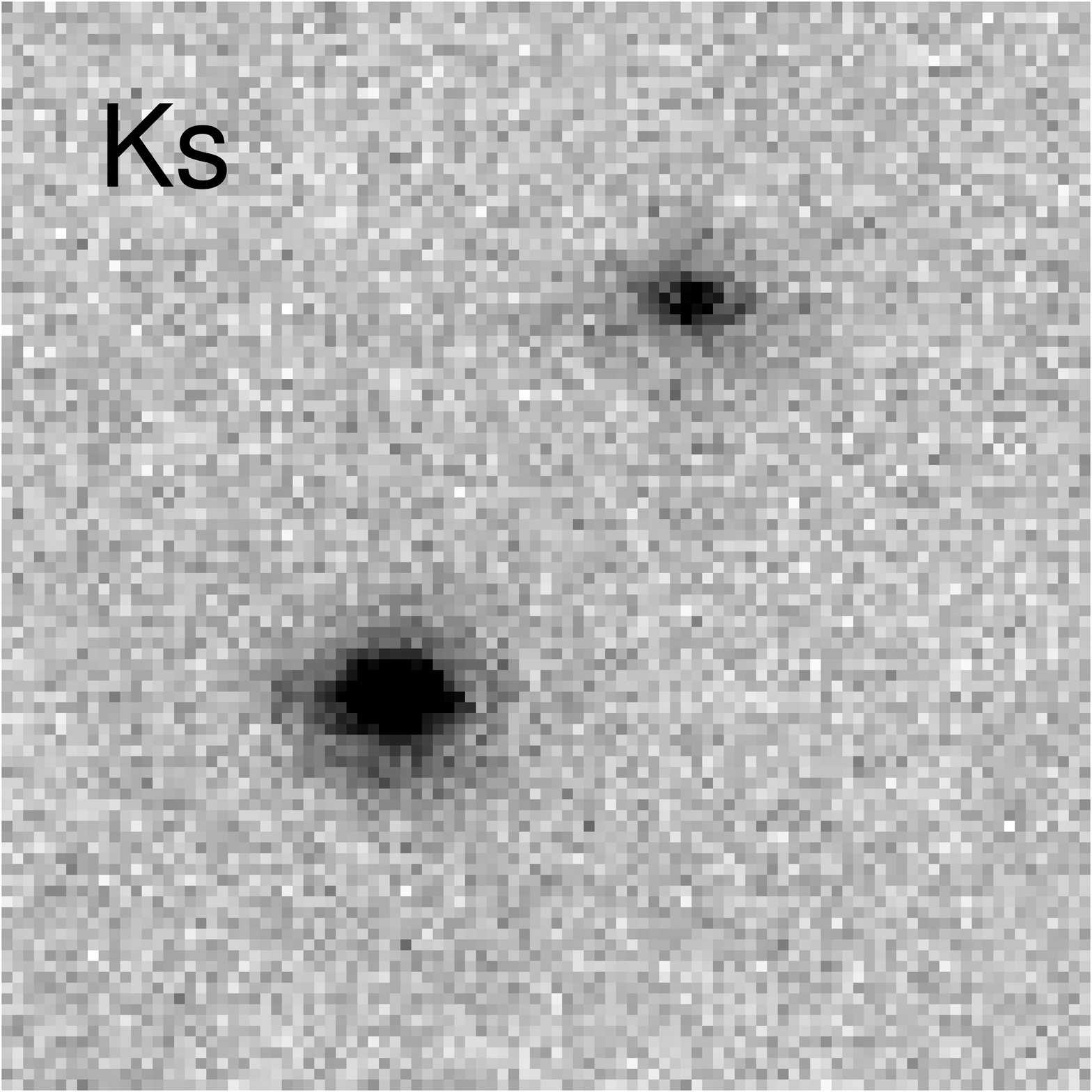}
\includegraphics[height=0.45\textwidth]{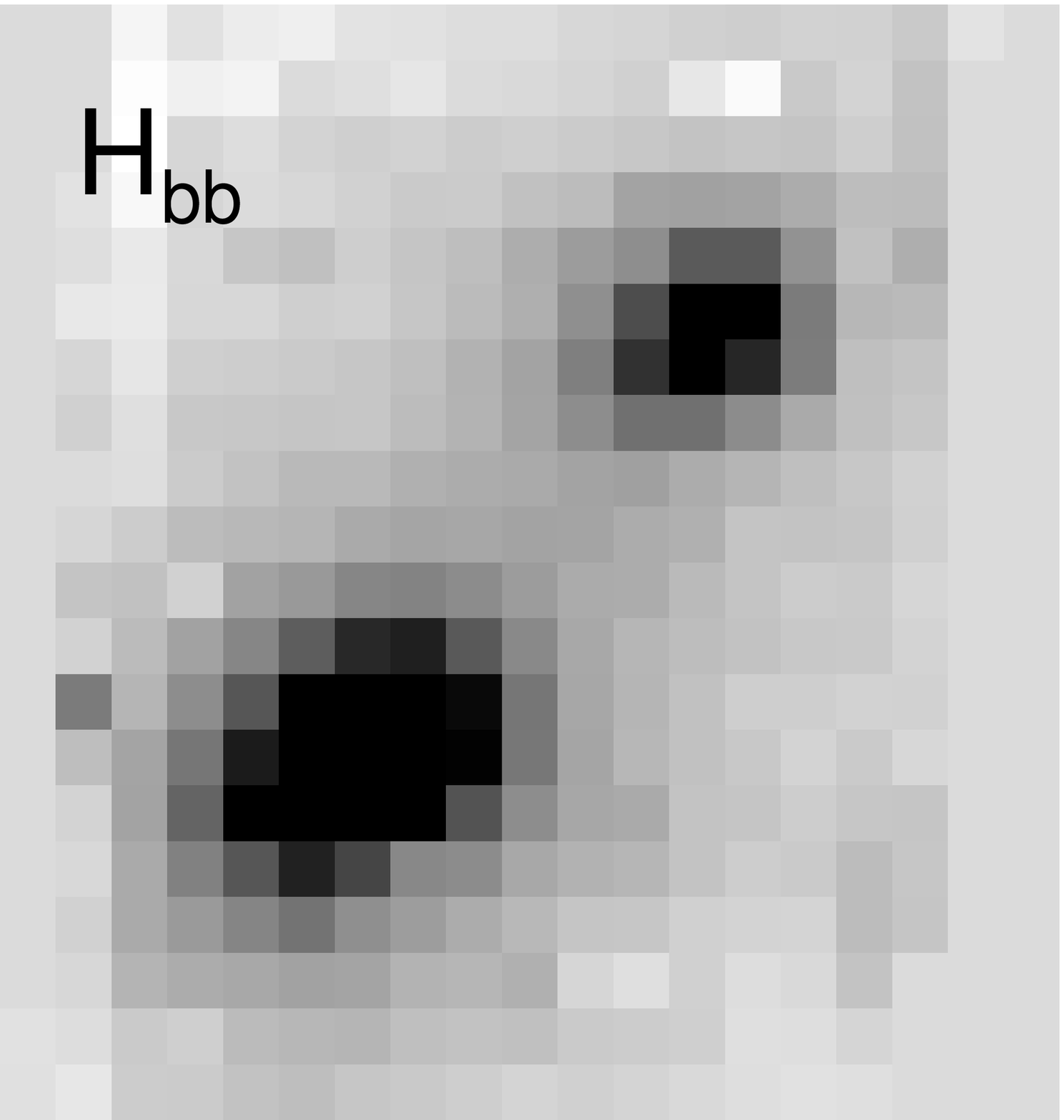}
\caption{Images of the {\namesh}AB system, from NIRC2 $J$- (top left), $H$- (top right) and $K_s$-band  (bottom left) observations obtained on 2011 August 29 (UT), and OSIRIS observations (bottom right) obtained on 2011 February 3 (UT).  The last image was generated by combining the OSIRIS data cube over wavelengths 1.56--1.60~$\micron$. All fields are 1$\arcsec$ on a side and oriented with North up and East to the left. 
\label{fig_image}}
\end{figure*}

Images were reduced using custom IDL\footnote{Interactive Data Language.} scripts. 
Sky background and dark current were removed from each image
by pairwise subtracting it with the following image in the dither sequence, 
then dividing by a normalized dome flat 
to correct for pixel-to-pixel sensitivity variations. The calibrated image frames
were registered to the peak of the primary's point spread function (PSF), then
median-combined to produce the final mosaics shown
in Figure~\ref{fig_image}.  

The quality of these data are significantly improved over the discovery observations from \citet{2011AJ....142...57G}, with components that are well-separated in the field-of-view (FOV).  
Relative fluxes were computed using aperture photometry on the coadded images.  Given the wide separation of the components, no PSF subtraction or other specialized procedures were required for the photometry.  The radial separation and position angle of the fainter component with respect to the brighter component was measured on each of the nine individual exposures.  The values listed in Table~\ref{tab_astrometry} report the average and standard deviation of those nine measurements. 
Relative $J$- and $H$-band flux measurements are in excellent agreement with those reported in 
\citet{2011AJ....142...57G}, but are of far greater precision.  In contrast, there is a significant shift in the relative separation of the two components, with $\Delta{\alpha}$ = 54$\pm$14~mas and $\Delta{\delta}$ = 27$\pm$16~mas.  Notably, this shift is largely in a radial direction.  We discuss the astrometry in further detail in Section~3.

\subsection{OSIRIS Spectroscopy}

Resolved spectroscopy of {\namesh} was obtained with OSIRIS and LGSAO  on 2011 February 3 (UT), in clear conditions with seeing of 0$\farcs$8-1$\farcs$0.   
The 50~mas-scale camera and H$_{bb}$ filter were employed, 
providing 1.47--1.80~$\micron$ spectroscopy at an average resolution of 3800 and dispersion of 2.1~{\AA}~pixel$^{-1}$ over a 0$\farcs$8$\times$3$\farcs$2 FOV.
The source was first acquired in the imaging camera, then offset to the spectrograph array. 
The image rotator was set to 0$\degr$.
Eight exposures of 600~s each were obtained over an airmass range of 1.40--1.52
using a two-step dither pattern with steps of 1$\farcs$0 along the long axis of the FOV.
For LGSAO tip-tilt correction, we employed the same star as that used for the NIRC2 imaging.
We also obtained four 30s observations of the A0~V star BD~+64~489 ($V$ = 9.43) in natural guide star AO mode at an airmass of 1.72 for telluric correction and flux calibration purposes.

Data were reduced with the OSIRIS data reduction pipeline \citep{2004SPIE.5492.1403K}, version 2.3.  
We first subtracted from the source and calibrator images median-combined sets of dark frames with identical
exposure times.  We then used the pipeline to 
adjust bias levels, remove detector artifacts and cosmic rays, extract and wavelength-calibrate the position-dependent spectra (using the most current rectification files as of February 2011), assemble 3D data cubes and correct for dispersion. Figure~\ref{fig_image} displays a 1$\arcsec$$\times$1$\arcsec$ section of this cube coadded 
over the wavelength range 1.56--1.60~$\micron$.   The two components are cleanly resolved, separated by $\sim$10~spaxels (0$\farcs$5) along a NW-SE axis.  The full width at half maximum of the PSF was measured to be 150~mas (3 spaxels) over this wavelength range for both components, 
slightly elongated in the direction of the tip-tilt star.
Relative astrometry from these data were determined from each of the reduced data cubes, using a median stack over the spectral dimension 1.56--1.60~$\micron$.  The IDL routine {\em cntrd} was used to measure the peak pixel positions of the primary and secondary in each image, and pixel separations along each axis were converted to angular separations in Right Ascension and declination. 
Mean measurements and standard deviations are listed in Table~\ref{tab_astrometry}.  
The offsets are intermediate between those of the two NIRC2 observations, consistent with a continuous reduction in separation over nearly constant position angle.

\begin{figure*}
\centering
\epsscale{1.0}
\plotone{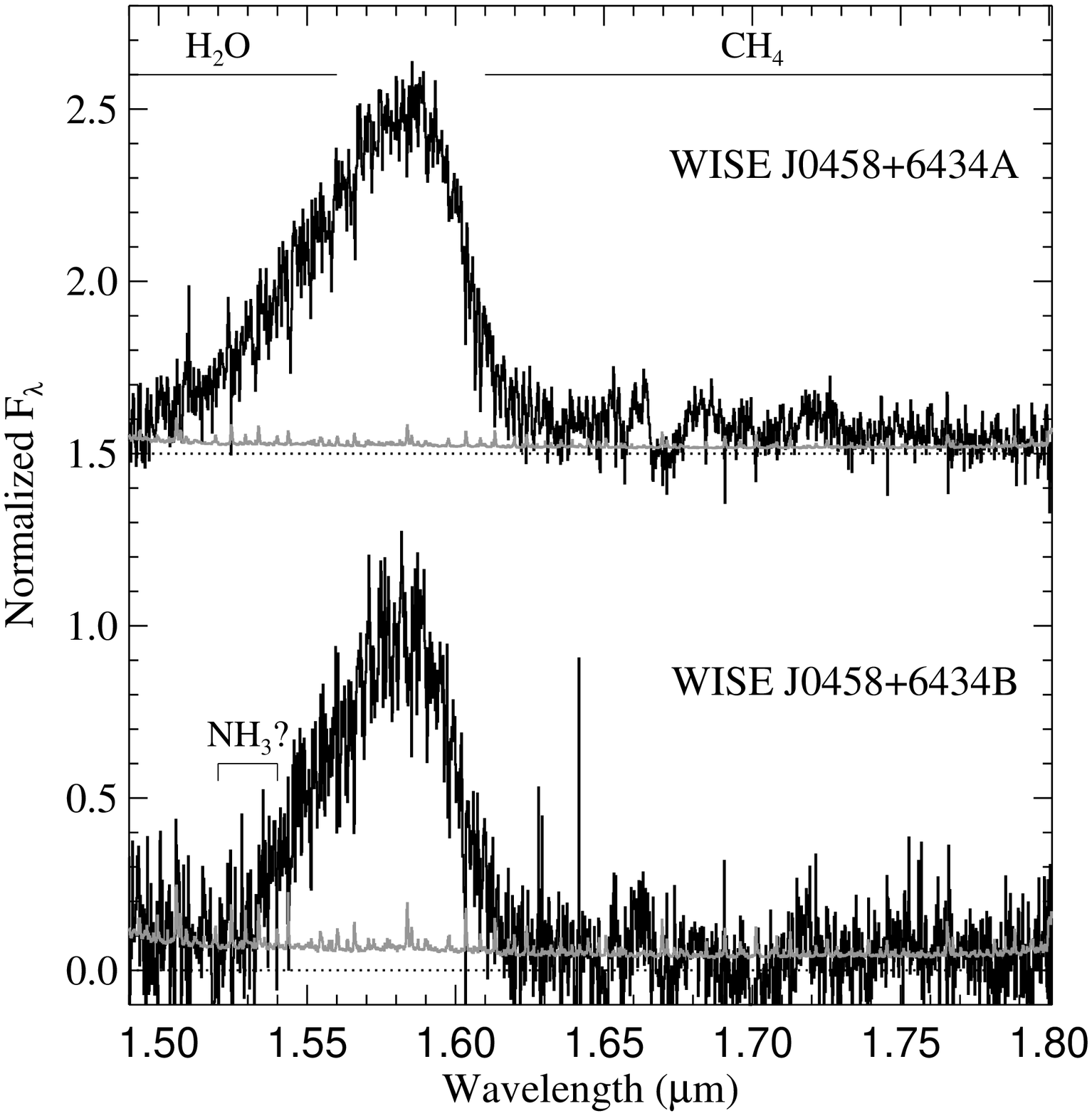}
\caption{Reduced OSIRIS spectra (black lines) of {\namesh}A (top) and B (bottom) over 1.5--1.8~$\micron$.
Spectra are normalized at 1.58~$\micron$, with data for {\namesh}A offset by 1.5 units for clarity (dotted lines).
Uncertainty spectra are indicated by grey lines. 
Regions of strong {\wat} and {\meth} absorption bands are labeled, as well as the 1.52-1.54~$\micron$ region of  {\ammon} absorption tentatively indicated in the spectrum of {\namesh}B.
\label{fig_spec}}
\end{figure*}

Spectra for both components of {\namesh} and the calibrator star 
were extracted directly from the data cube via aperture photometry in each image plane, using a 1.5-spaxel (75~mas) aperture for {\namesh} and a 3-spaxel (150~mas) aperture for  BD~+64~489, with
10--20~spaxel (0$\farcs$5--1$\farcs$0) sky annuli.  
The individual spectra for all three sources were scaled and combined using the {\it xcombspec} routine in SpeXtool \citep{2004PASP..116..362C}.  Flux calibration and telluric correction of the {\namesh}AB spectra were performed using the {\it xtellcor\_general} routine in SpeXtool, assuming a 20~nm Gaussian kernel to model the widths of the 
A0V H~I lines, and a Kurucz model spectrum of Vega \citep{2003PASP..115..389V}.
Figure~\ref{fig_spec} displays the reduced spectra of the {\namesh} components.
Both are unambiguously late-type T dwarfs based on the presence of strong {\wat} absorption up to 1.55~$\micron$ and strong
{\meth} absorption beyond 1.59~$\micron$.  These bands constrain the $H$-band flux to a discrete 1.50--1.62~$\micron$ peak, which is narrower in the spectrum of {\namesh}B.  Unbinned peak signal-to-noise (S/N) is $\sim$40 in the spectrum of {\namesh}A and $\sim$15 in the spectrum of {\namesh}B.

\section{Astrometric Analysis}

\begin{figure}
\centering
\epsscale{0.8}
\plotone{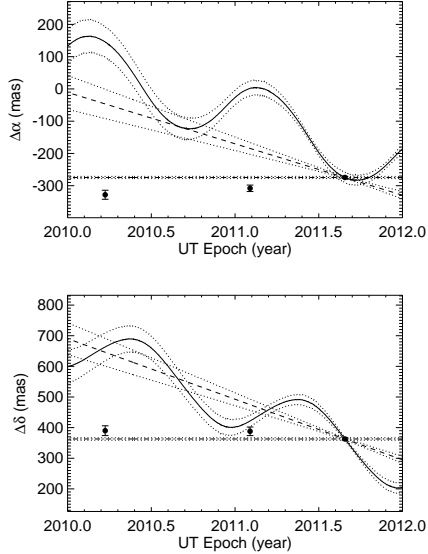}
\caption{Separation of the {\namesh} components in Right Ascension (top) and declination (bottom) as a function of time, as measured from primary to secondary.  Measurements from NIRC2 and OSIRIS are indicated by points with error bars.  Lines delineate predicted separations relative to epoch 2011 August 29 (UT) astrometry assuming (1) the secondary is a distant background source (solid line), (2) the secondary is an unmoving source at the same distance as the primary (dashed line) and (3) the separation between primary and secondary remains constant.  Dotted lines indicate 1$\sigma$ uncertainties in these trajectories based on uncertainties in the system's distance, total proper motion and separation at epoch 2011 August 29 (UT).
\label{fig_astrometry}}
\end{figure}

As noted above, astrometry from NIRC2 and OSIRIS observations of {\namesh} indicate a monotonic decrease in the separation of this pair along a near constant position angle of 320$\degr$, with the separations in Right Ascension from the two NIRC2 observations differing by nearby 4$\sigma$.
To assess how this relative motion relates to systemic common motion, and hence physical association, we compared our measurements against predicted separations anchored to our 2011 August 29 (UT) astrometry, considering three limiting scenarios: (1) the secondary is a background source with negligible parallactic or proper motion, (2) the secondary is at the same distance as the primary, but with negligible proper motion, and (3) the primary and secondary maintain constant separation at all times.  Figure~\ref{fig_astrometry} shows that none of these scenarios are fully consistent with the data, although the last provides the closest match.  Allowing relative motions and distances to vary freely, we found that our measured astrometry (with 3$\sigma$ uncertainty allowance) constrains the relative proper motion to be less than 77~mas~yr$^{-1}$, or a mere 4~{\kms} at a distance of 10~pc.  For comparison, the circular orbital speed of a 0.05~{\msun} brown dwarf in an equal-mass binary separated by 5~AU is $\sim$3~{\kms}.  We also measured the relative radial velocity ($\Delta$RV) of the two components by cross-correlating our OSIRIS spectra over the 1.58--1.65~$\micron$ range, properly accounting for noise through Monte Carlo simulation.  We find $\Delta$RV = 3$\pm$15~{\kms}, consistent with zero to within 1/5 of a spectral resolution element.

\begin{figure}
\centering
\epsscale{0.8}
\plotone{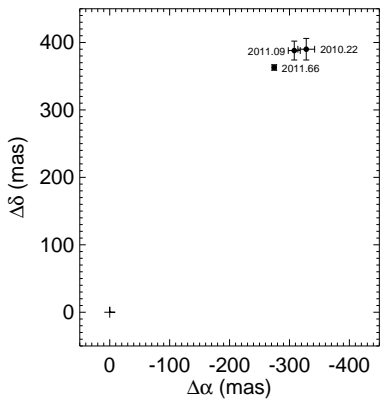}
\caption{Relative separations of {\namesh}A (cross) and B (data points) projected onto the plane of the sky. The trend in motion is nearly radial, toward a smaller separation.
\label{fig_orbit}}
\end{figure}

\begin{deluxetable}{lc}
\tabletypesize{\small}
\tablecaption{Astrometry for {\namesh}AB\label{tab_astrometry}}
\tablewidth{0pt}
\tablehead{
\colhead{Parameter} &
\colhead{Value} 
}
\startdata
\cline{1-2}
\multicolumn{2}{c}{NIRC2 Epoch 2010 March 24 (UT)\tablenotemark{a}} \\
\cline{1-2}
$\Delta{\alpha}\cos{\delta}$ (mas) & $-$328$\pm$14    \\
$\Delta{\delta}$ (mas) & 390$\pm$16 \\
$\rho$ (mas) & 510$\pm$20   \\
$\theta$ ($\deg$) & 320$\pm$1   \\
$\Delta{J}$ & 0.98$\pm$0.08 \\
$\Delta{H}$ & 1.00$\pm$0.09 \\
\cline{1-2}
\multicolumn{2}{c}{OSIRIS Epoch 2011 February 3 (UT)} \\
\cline{1-2}
$\Delta{\alpha}\cos{\delta}$ (mas) & $-$308$\pm$10   \\
$\Delta{\delta}$ (mas) & 386$\pm$14   \\
$\rho$ (mas) & 493$\pm$15   \\
$\theta$ ($\deg$) & 321.4$\pm$1.0   \\
\cline{1-2}
\multicolumn{2}{c}{NIRC2 Epoch 2011 August 29 (UT)} \\
\cline{1-2}
$\Delta{\alpha}\cos{\delta}$ (mas) & $-$274.5$\pm$3.4    \\
$\Delta{\delta}$ (mas) & 362.9$\pm$4.0 \\
$\rho$ (mas) & 455.1$\pm$4.2   \\
$\theta$ ($\deg$) & 322.9$\pm$0.4   \\
$\Delta{J}$ & 0.98$\pm$0.01 \\
$\Delta{H}$ & 1.02$\pm$0.01 \\
$\Delta{K_s}$ & 1.06$\pm$0.03 \\
\enddata
\tablenotetext{a}{From \citet{2011AJ....142...57G}.}
\tablecomments{Angular separation ($\rho$) and position angle ($\theta$) are measured from the brighter primary to the fainter secondary.}
\end{deluxetable}

These stringent limits on the relative motion of the {\namesh} components indicate that the two sources comprise a common proper motion, physically-bound binary.  As such, the small but significant astrometric shifts observed must arise from orbital motion.  Figure~\ref{fig_orbit} provides a visualization of our three separation measurements projected onto the plane of the sky. The near radial motion and marginal indication of an inward acceleration ($\ddot{\rho} = -1.6{\pm}1.3$~{\kms}~yr$^{-1}$) suggests an eccentric and/or nearly edge-on orbit that is moving toward alignment.  In the most optimistic case---if {\namesh} was observed at or near maximum elongation in the first NIRC2 epoch---these observations constrain the orbital semimajor axis to the range 5/(1+e) $\lesssim a \lesssim$ 5/(1$-$e)~AU, where $e$ is the orbital eccentricity and a distance of 10~pc is assumed.  Unfortunately, more stringent constraints on this system's orbit are not possible with these few measurements clumped in one region of separation space.  Over the next decade it should be possible to disentangle the eccentricity and inclination of the orbit, and provide more robust constraints on the period and total system mass.

\section{Spectral Analysis}

\subsection{Component Spectral Classifications}

\begin{figure*}
\centering
\epsscale{1.0}
\plottwo{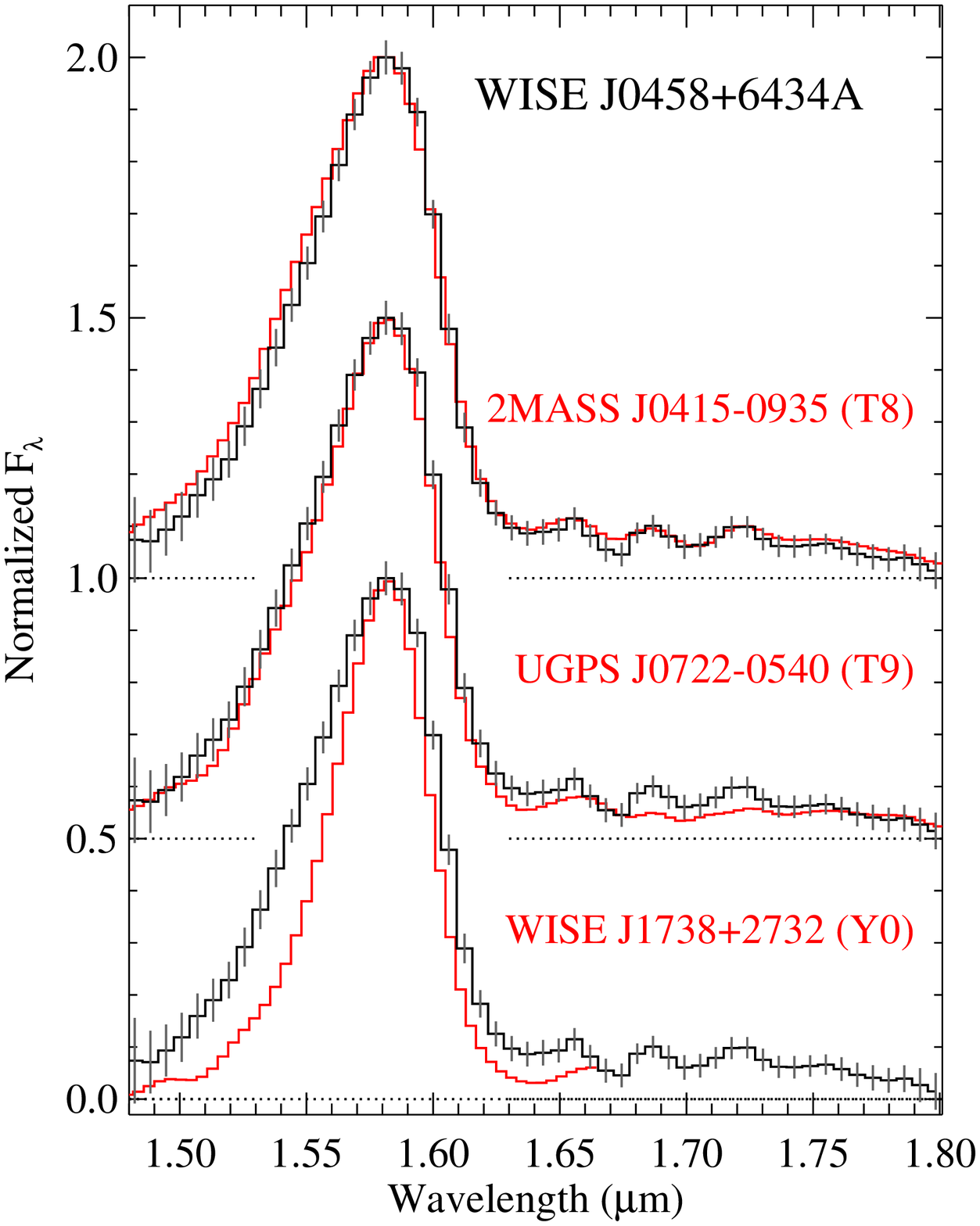}{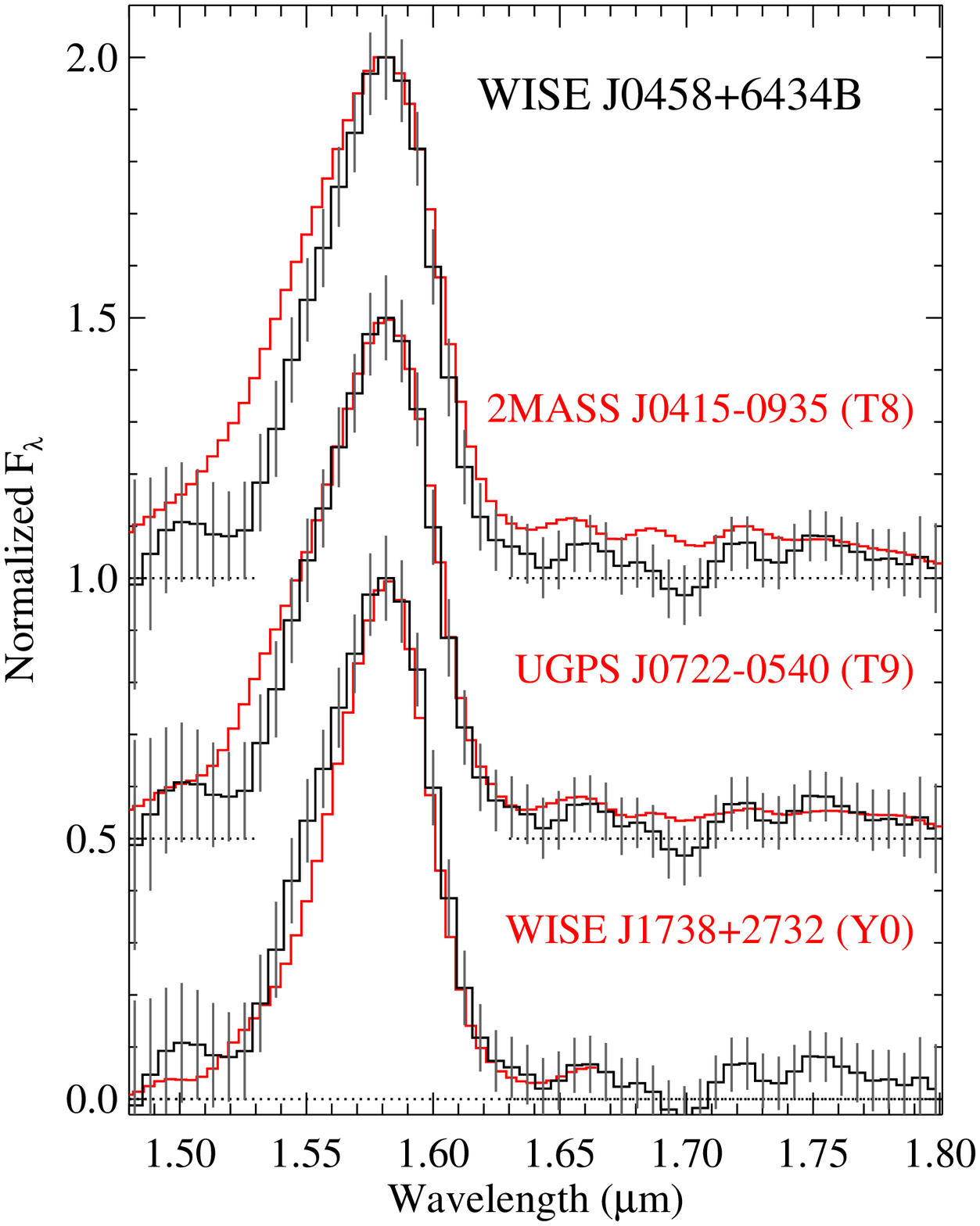}
\caption{Boxcar-smoothed OSIRIS spectra (black lines) of {\namesh}A (left) and {\namesh}B (right) compared to spectral standards (red lines) 2MASS~J0415$-$0935 (T8, top), UGPS~J0722$-$0540 (T9; middle)
and WISE~J1738+2732 (Y0, bottom).  All spectra are smoothed to a common resolution of {\ldl} = 100
and normalized at 1.58~$\micron$.  Uncertainties in the smoothed {\namesh}AB spectra are indicated by error bars; note that these do not take into account correlated errors.  Zeropoints are indicated by dotted lines.
\label{fig_classification}}
\end{figure*}

\begin{deluxetable*}{llllcl}
\tabletypesize{\footnotesize}
\tablecaption{Spectral Indices and Classification \label{tab_indices}}
\tablewidth{0pt}
\tablehead{
\colhead{Source} &
\colhead{{\wat}-H\tablenotemark{a}} &
\colhead{{\meth}-H\tablenotemark{a}} &
\colhead{{\ammon}-H\tablenotemark{a}} &
\colhead{Adopted} &
\colhead{Ref} \\
 & & & & \colhead{SpT\tablenotemark{b}} \\
}
\startdata
2MASS~J0415$-$0935\tablenotemark{c} & 0.183$\pm$0.002 & 0.104$\pm$0.002 & 0.628$\pm$0.004 &  T8 & 1,2 \\
{\bf {\namesh}A} & 0.129$\pm$0.006 (T9) & 0.079$\pm$0.003 (T9) & 0.561$\pm$0.003 (T8.5)  & T8.5 & 3 \\
UGPS~J0722$-$0540\tablenotemark{c} & 0.115$\pm$0.006 & 0.075$\pm$0.005 & 0.527$\pm$0.008 & T9 & 4,5 \\
{\bf {\namesh}B} & 0.086$\pm$0.015 (T9.5) & 0.040$\pm$0.007 (Y0+) & 0.459$\pm$0.008 (T9.5) & T9.5 & 3 \\
WISE~J1738+2732\tablenotemark{c} & 0.045$\pm$0.008 & 0.052$\pm$0.008 & 0.350$\pm$0.011 & Y0 & 5 \\
\enddata
\tablenotetext{a}{Index uncertainties computed directly from spectral flux errors via Monte Carlo sampling.}
\tablenotetext{b}{For {\namesh}AB, this includes comparison to spectral standards as shown in Figure~\ref{fig_classification}.}
\tablenotetext{c}{Spectroscopic standards \citep{2006ApJ...637.1067B,2011arXiv1108.4678C}.}
\tablerefs{(1) \citet{2002ApJ...564..421B}; (2) \citet{2006ApJ...637.1067B}; (3) This paper; (4) \citet{2010MNRAS.408L..56L}; (5) \citet{2011arXiv1108.4678C}.}
\end{deluxetable*}

The strong {\wat} and {\meth} absorption in the spectra of {\namesh}AB 
are consistent with very late T and possibly Y spectral types. 
To robustly determine the classifications, we first compared the
spectra to the standards 
2MASS~J04151954$-$0935066 (T8; hereafter 2MASS~J0415$-$0935; \citealt{2002ApJ...564..421B,2004AJ....127.2856B}),
UGPS~J072227.51$-$054031.2 (T9; hereafter UGPS~J0722$-$0540; \citealt{2010MNRAS.408L..56L,2011arXiv1108.4678C}),
and WISEP J173835.52+273258.9 (Y0; hereafter WISE~J1738+2732; \citealt{2011arXiv1108.4678C}).
The latter two sources were defined by \citet{2011arXiv1108.4678C} to extend the T dwarf near-infrared classification scheme of 
\citet{2006ApJ...637.1067B} across the T dwarf/Y dwarf transition.
Figure~\ref{fig_classification} shows these comparisons, with all of the spectra
smoothed to a common resolution of {\ldl} = 100 
and normalized at 1.58~$\micron$.  
{\namesh}A is a reasonably fair match to 
2MASS~J0415$-$0935, albeit with slightly deeper absorption on the blue side of the 1.58~$\micron$ peak and a somewhat narrower peak.  It has weaker 1.6~$\micron$ {\meth} absorption than UGPS~J0722$-$0540, indicating that it is not as late as T9.  {\namesh}B, on the other hand, has stronger {\wat} and {\meth} absorption than 2MASS~J0415$-$0935 and UGPS~J0722$-$0540, but less absorption around 1.55~$\micron$ 
than WISE~J1738+2732.  These by-eye comparisons suggest types of T8-T8.5 for {\namesh}A and T9-T9.5 for {\namesh}B.

We also computed $H$-band spectral classification indices {\wat}-H and {\meth}-H  \citep{2006ApJ...637.1067B} and  {\ammon}-H  \citep{2008A&A...482..961D} for both components of {\namesh}.  The 
last index was shown by \citet{2011arXiv1108.4678C} to exhibit a distinct break at the T/Y transition.
Values are listed in Table~\ref{tab_indices}, along with estimated subtypes based on comparison to index values for the spectral standards.  Indices for {\namesh}A are well-matched to those for  UGPS~J0722$-$0540, albeit with an {\ammon}-H index intermediate between T8 and T9.
{\wat}-H and {\ammon}-H indices for {\namesh}B are intermediate between T9 and Y0; its {\meth}-H index is smaller than any late T dwarf/Y dwarf measured thus far with the exception of the T9.5 
WISEPC~J014807.25$-$720258.8 \citep{2011arXiv1108.4678C}.  Combining the index types and direct spectral comparisons, we infer  subtypes of T8.5 and T9.5 for {\namesh}A and B.

\subsection{Ammonia in the Spectrum of {\namesh}B}

\begin{figure*}
\centering
\epsscale{1.1}
\plottwo{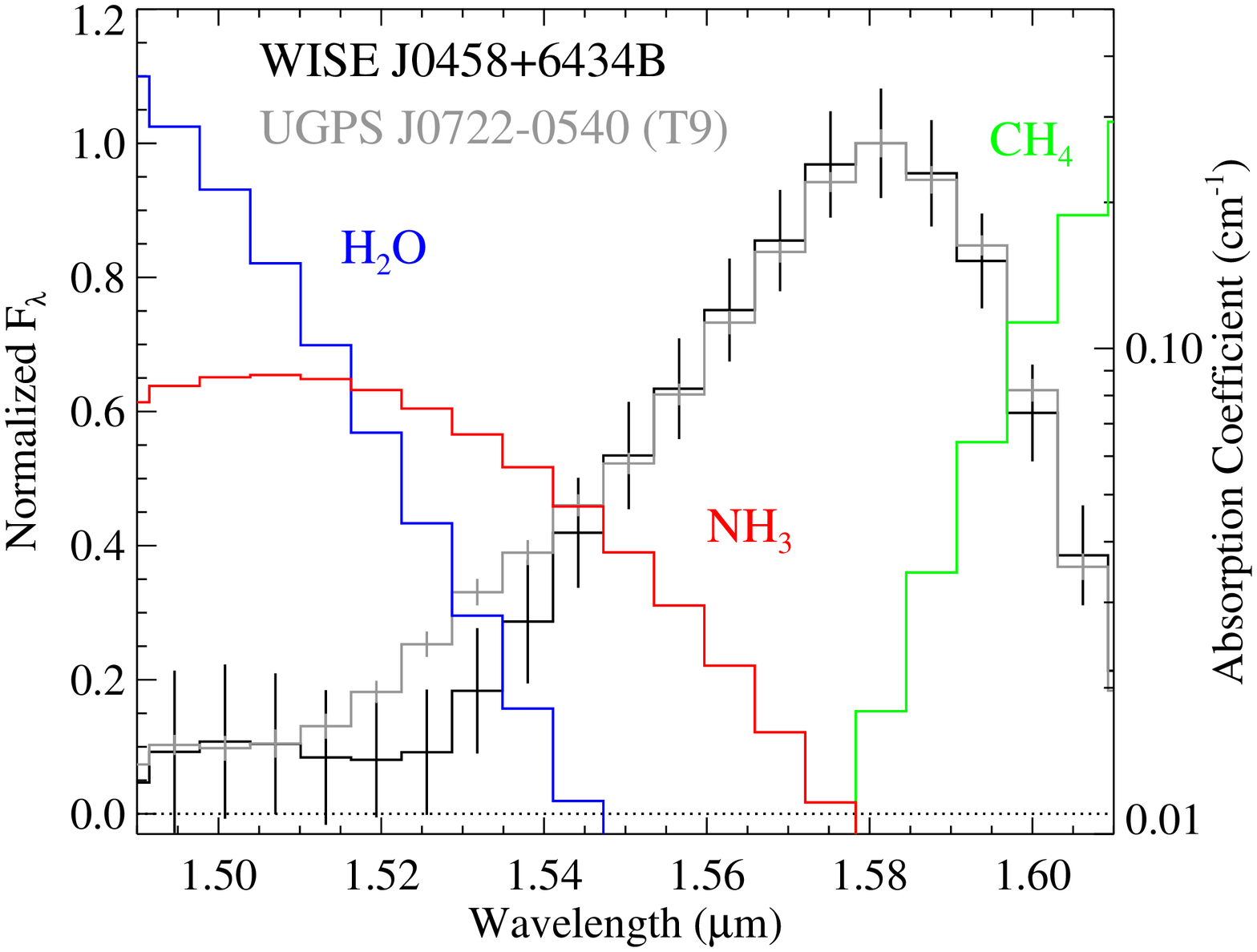}{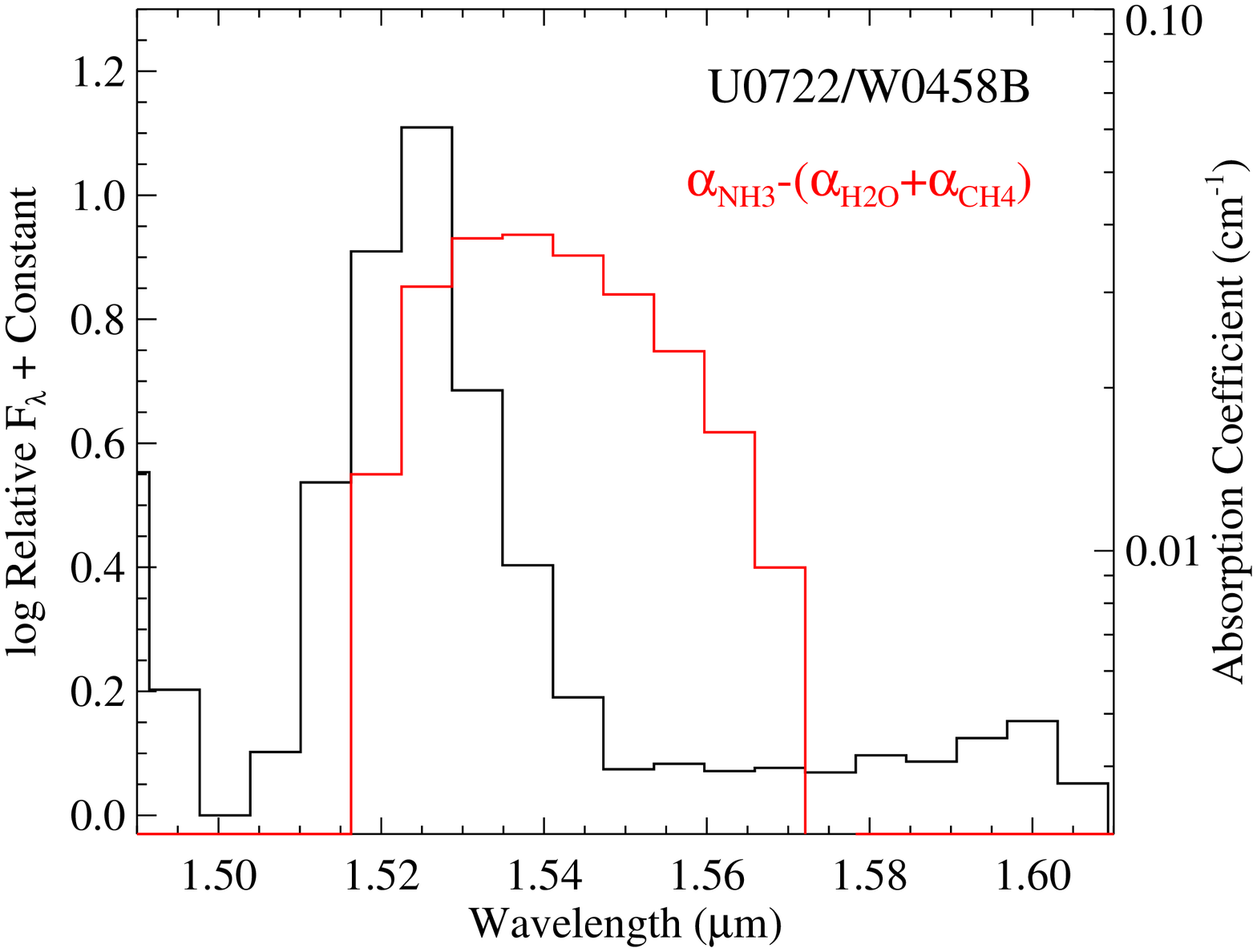}
\caption{(Left) 1.49--1.61~$\micron$ spectrum of {\namesh}B (black line) and UGPS~J0722$-$0540 (grey line), both smoothed to a common resolution of {\ldl} = 100 (uncertainties are indicated by error bars).  Overplotted are wavelength-dependent absorption coefficients for {\wat} (blue), {\meth} (green) and {\ammon} (red), based on scattering cross-sections from \citet{2008ApJS..174..504F} at T = 500~K and P = 1~bar, and assuming nonequlibrium abundances  from \citet{2006ApJ...647..552S}.  
Note the region longward of 1.52~$\micron$ where {\ammon} opacity is predicted to dominate over {\wat}.
(Right) Relative flux between UGPS~J0722$-$0540 and {\namesh}B (black line) on a logarithmic scale, compared to the relative absorption coefficient between {\ammon} versus {\wat} and {\meth} combined (red line).  The peak in relative flux is near the peak in relative absorption.
\label{fig_ammonia}}
\end{figure*}

Close examination of the spectral comparisons in Figure~\ref{fig_classification} indicates additional absorption in the spectrum of {\namesh}B over the 1.52--1.54~$\micron$ region compared to both
2MASS~J0415$-$0935 and UGPS~J0722$-$0540. This feature is similar to those noted by \citet{2011arXiv1108.4678C} in the spectra of the Y0 dwarfs WISEPC~J140518.40+553421.5 and WISE~J1738+2732, which were tentatively attributed to {\ammon} absorption.  We examined whether 
the excess absorption in {\namesh}B could also be due to {\ammon} by comparing its spectrum and that of UGPS~J0722$-$0540 to absorption coefficients ($\alpha[\lambda]$) for {\wat}, {\meth} and {\ammon}.  Absorption coefficients were computed by multiplying scattering cross-sections per molecule ($\sigma[\lambda]$), as tabulated by \citet{2008ApJS..174..504F} for T = 500~K and P = 1~bar, with molecular number densities ($n_i$) assuming fractional abundances ($f_i$) based on the non-equilibrium chemistry calculations of \citet{2006ApJ...647..552S}:
\begin{eqnarray*}
\alpha_i[\lambda] & = & \sigma_i[\lambda]n_i \\
 & = & \sigma_i[\lambda]f_in.
\end{eqnarray*}
Here, $n = 1.4\times10^{19}$~cm$^{-3}$ is the total number density of gas particles assuming an ideal gas.  From \citet{2006ApJ...647..552S}, we adopt $\log_{10}f_{H_2O}$ = $-$3.1, $\log_{10}f_{CH_4}$ = $-$3.3 and $\log_{10}f_{NH_3}$ = $-$4.9.  Note that non-equilibrium chemistry reduces  {\ammon} abundances by a factor of 10.

Figure~\ref{fig_ammonia} compares the smoothed absorption coefficients for these molecules to equivalently smoothed spectra of UGPS~J0722$-$0540 and {\namesh}B.  We find that {\ammon} absorption exceeds that of {\wat} beyond 1.52~$\micron$, around the region where we also see excess absorption in the spectrum of {\namesh}B.  The right panel of Figure~\ref{fig_ammonia} shows another view of this, comparing the relative absorption between UGPS~J0722$-$0540 and {\namesh}B to the differential absorption between {\ammon}, {\wat} and {\meth}.  
There is relatively close spectral alignment between excess {\ammon} opacity and excess absorption in the spectrum of {\namesh}B, suggesting {\ammon} as the primary absorber at these wavelengths.
However, this conclusion carries a few caveats. First, the photospheres of brown dwarfs are not homogeneous, and absorption contributing to this spectral region is integrated over a pathlength that spans a range of temperatures and pressures that do not necessary coincide with the values used here to calculate molecular opacities.  Second, the opacities of absorbers other than {\ammon} are also likely to differ between these two sources, which may add additional structure to the relative intensities.  Finally, while the excess absorption in {\namesh}B is statistically significant, the precise location of the relative intensity peak is not robust due to the large uncertainties in the spectrum of {\namesh}B at these (low intensity) wavelengths.  Hence, while evidence for {\ammon} absorption in the near-infrared spectrum of {\namesh}B is compelling, confirmation of this will require higher signal-to-noise data and more detailed modeling
of the spectrum.

\section{Discussion}

\begin{figure*}[h]
\centering
\includegraphics[width=0.75\textwidth]{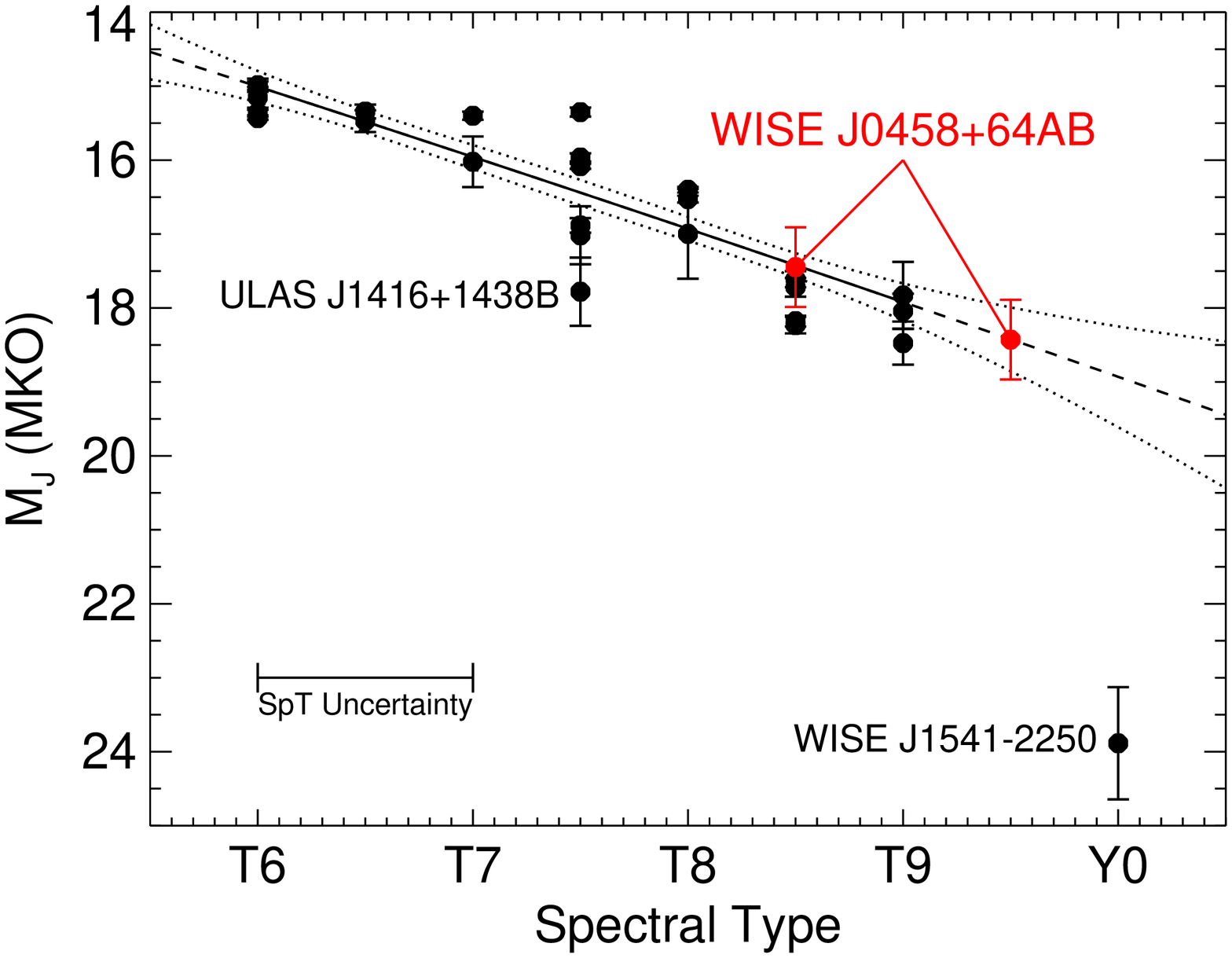}
\includegraphics[width=0.75\textwidth]{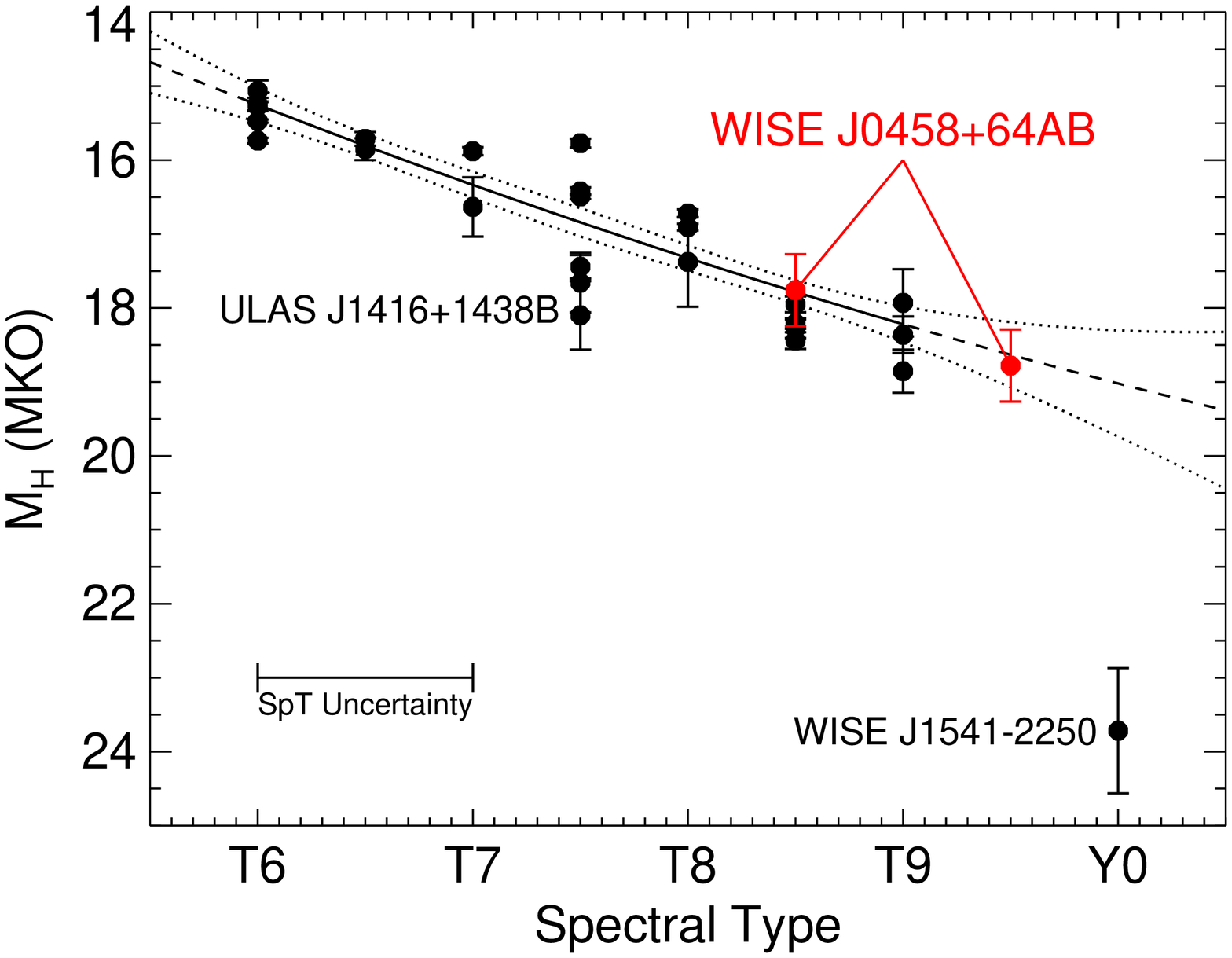}
\caption{Absolute MKO $J$- (top) and $H$-band (bottom) magnitudes versus spectral type for 27 T6--Y0 dwarfs with measured parallaxes. Photometric uncertainties are indicated; spectral type uncertainties are assumed to be $\pm$0.5 subtypes.  Second-order polynomial fits to the T6--T9 dwarfs based on Monte Carlo uncertainty sampling yield a mean relation indicated by the solid line (dashed for extrapolation beyond T9) and 1$\sigma$ uncertainties indicated by the dotted lines.  The estimated absolute magnitudes and uncertainties for {\namesh}A based on this relation are indicated, as well as those for {\namesh}B assuming it lies at the same distance as the A component.   
\label{fig_absmag}}
\end{figure*}

One of the benefits of studying a physical binary is that its components share
a common distance.  In the case of {\namesh}AB, this allows us to bootstrap existing absolute magnitude/spectral type trends out to the end of the T dwarf sequence.  Figure~\ref{fig_absmag} shows absolute MKO $J$ and $H$ magnitudes for 27 unresolved T6-Y0 dwarfs with reported parallax distance measurements\footnote{These are based on data from \citet{2003AJ....126..975T, 2004A&A...413.1029M, 2004AJ....127.2948V, 2006AJ....132.2360H, 2009MNRAS.395.1237B, 2010ApJ...718L..38A, 2010MNRAS.405.1140G, 2010ApJ...710.1627L, 2010A&A...510L...8S, 2010MNRAS.408L..56L, 2010A&A...524A..38M, 2011arXiv1108.4677K} and \citet{2011arXiv1103.0014L}. We excluded two sources from this sample: the T6.5 SDSSp~J134646.45$-$003150.4, for which existing photometry is blended with an unassociated background source (Burgasser et al., in prep.); and the T6.5p ULAS~J115038.79+094942, which has a poorly constrained parallax ($\sigma_{\pi}/{\pi}$ = 0.45; \citealt{2010A&A...524A..38M}).}.  For sources with types later than T8, we adopt the revised classifications listed in \citet{2011arXiv1108.4678C}.
To delineate trends in absolute magnitudes from T6 to T9, we generated a series of second-order polynomial fits to these data accounting for uncertainties through Monte Carlo sampling\footnote{We assumed classification uncertainties of 0.5~subtypes and absolute magnitude uncertainties as indicated.  We then performed 1000 fits to simulated data, offsetting the original spectral types and absolute magnitudes of each source by random draws from a normal distribution with mean zero and standard deviation equal to the uncertainty.}.  Figure~\ref{fig_absmag} shows the resulting mean trends and 1$\sigma$ deviations in $M_J$ and $M_H$ as a function of spectral type, which were used to estimate the absolute magnitudes for {\namesh}A and B (Table~\ref{tab_parameters}).  Combining absolute and apparent magnitudes for each component in both filter bands, we derive statistically consistent distance estimates of 10.5$\pm$1.8~pc and 11.2$\pm$2.2~pc for the A and B components, respectively.  These are also formally consistent with the systemic spectrophotometric estimates of \citet[9.0$\pm$1.9~pc]{2011ApJ...726...30M} and \citet[10.5$\pm$1.4~pc]{2011AJ....142...57G}.  The agreement between these distances further supports the identification of this system as a physical binary; or, assuming that, justifies the extrapolation of existing absolute magnitude/spectral type trends to subtype T9.5.

\begin{deluxetable}{lcc}
\tabletypesize{\footnotesize}
\tablecaption{Component Photometry and Distances for {\namesh}AB \label{tab_parameters}}
\tablewidth{0pt}
\tablehead{
\colhead{Parameter} &
\colhead{{\namesh}A} &
\colhead{{\namesh}B} \\
}
\startdata
MKO $J$ & 17.50$\pm$0.07 & 18.48$\pm$0.07   \\
MKO $M_J$ & 17.4$\pm$0.5 & 18.4$\pm$0.5  \\
MKO $H$ & 17.77$\pm$0.11 & 18.79$\pm$0.11   \\
MKO $M_H$ & 17.8$\pm$0.5 & 18.8$\pm$0.5   \\
Est.~Distance (pc) & 10.5$\pm$1.8 & 11.2$\pm$2.2  \\
\enddata
\tablecomments{Component brightnesses are based on the new NIRC2 relative photometry reported here, combined light photometry from \citet{2011ApJ...726...30M}, and the 2MASS-MKO filter corrections given in \citet{2011AJ....142...57G}.   Absolute magnitudes are based on the relations shown in Figure~\ref{fig_absmag}.}
\end{deluxetable}

Currently, the only Y dwarf  with a parallax distance measurement is WISEP~J154151.65$-$225025.2 (hereafter WISE~J1541$-$2250), classified Y0 by \citet{2011arXiv1108.4678C}.  \cite{2011arXiv1108.4677K} report a preliminary parallax of 351$\pm$108~mas for this source, corresponding to a distance of 2.8$\pm$0.9~pc and absolute magnitudes $M_J$ = 23.9$\pm$0.8 and $M_H$ = 23.7$\pm$0.8.  These values make WISE~J1541$-$2250 roughly 100 times fainter than {\namesh}B in the near-infrared, despite the seemingly minor change in spectral morphology (cf.\ comparison of {\namesh}B to the Y0 standard WISE~J1738+2732 in Figure~\ref{fig_classification}).
This jump in brightness, also discussed by \citet{2011arXiv1108.4678C}, suggests a dramatic redistribution of radiation out of the near-infrared band, as the corresponding difference in {\teff}, roughly 500~K to 350~K, reduces the luminosity by only a factor of 4, assuming equivalent radii.  A greater fraction of flux is certainly being emitted at mid-infrared wavelengths, due to the redward shift in the blackbody peak and increased near-infrared absorption from {\wat}, {\meth} and collision-induced H$_2$ with lower {\teff}. Additional absorption may also arise from water ice clouds that are expected to form at these temperatures, as scattering opacity from 10-100~$\micron$-sized ice grains would efficiently suppress near-infrared emission just as mineral grains are observed to do so in the L dwarfs \citep{2001ApJ...556..872A}.  While an improved parallax measurement for WISE~J1541$-$2250 and other newly-discovered Y dwarfs are needed to confirm this dramatic drop in near-infrared flux, these early indications suggest that the T/Y transition, like other late spectral class transitions, may trace significant changes in atmospheric chemistry as well as declining temperatures.

\acknowledgments

The authors would like to thank Keck observing assistants Terry Stickel and Cynthia Wilburn, and instrument scientists Scott Dahm and Hien Tran, for their assistance in the observations reported here.  We also acknowledge guidance on OSIRIS data reduction from Breann Sitarski and Shelley Wright, and Richard Freedman for providing scattering cross section spectra for various molecules.
We thank our referee for her/his helpful and prompt review.
This work was supported by a NASA Keck PI Data Award, administered by the NASA Exoplanet Science Institute.
This research has made use of the SIMBAD database,
operated at CDS, Strasbourg, France;  
the M, L, and T dwarf compendium housed at \url{http://dwarfarchives.org};
and the SpeX Prism Spectral Libraries at \url{http://www.browndwarfs.org/spexprism}.
The authors wish to recognize and acknowledge the 
very significant cultural role and reverence that 
the summit of Mauna Kea has always had within the 
indigenous Hawaiian community.  We are most fortunate 
to have the opportunity to conduct observations from this mountain.

Facilities: \facility{Keck II (NIRC2, OSIRIS, LGSAO)}

\end{document}